\documentclass[12pt, reqno]{article}

\usepackage{latexsym}
\usepackage{amsthm}
\usepackage{amsmath}
\usepackage{epsfig}
\usepackage{amssymb}
\usepackage{amsbsy}
\usepackage{color}
\usepackage{mathrsfs}

\renewcommand{\d}{\delta}

\renewcommand{\d}{\partial}
\newcommand{\be}{\begin{eqnarray}}
\newcommand{\ee}{\end{eqnarray}}

\definecolor{downstairs}{rgb}{0.09, 0.1328, 0.6388}
\definecolor{upstairs}{rgb}{0.698,0.1259,0.259}

\makeindex
\begin{document}

\baselineskip=18pt

\setcounter{footnote}{0}
\setcounter{figure}{0}
\setcounter{table}{0}

\begin{titlepage}

\begin{center}

{\Large \bf Resultants and Gravity Amplitudes}

\vspace{0.5cm}

{\bf Freddy Cachazo}

\vspace{.1cm}

{\it Perimeter Institute for Theoretical Physics, Waterloo, Ontario N2J W29, CA}

\end{center}

\vspace{0.5cm}

\begin{abstract}

Two very different formulations of the tree-level S-matrix of ${\cal N}=8$ Einstein supergravity in terms of rational maps are known to exist. In both formulations, the computation of a scattering amplitude of $n$ particles in the $k$ R-charge sector involves an integral over the moduli space of certain holomorphic maps of degree $d=k-1$. In this paper we show that both formulations can be simplified when written in a manifestly parity invariant form as integrals over holomorphic maps of bi-degree $(d,\tilde d)$ with $\tilde d=n-d-2$. A map $\sigma \to (\lambda(\sigma ),\tilde\lambda(\sigma ))$ of bi-degree $(d,{\tilde d})$ is constructed using two maps from ${\mathbb CP}^1\to {\mathbb CP}^1$ denoted by $\lambda(\sigma) = (\lambda_1(\sigma),\lambda_2(\sigma))$ and $\tilde\lambda(\sigma) = (\tilde\lambda_1(\sigma),\tilde\lambda_2(\sigma ))$. In one formulation the full integrand becomes quite directly the product of the resultants of the polynomials defining each map, i.e, ${\rm R}(\lambda ){\rm R}(\tilde\lambda )$. In the second formulation, a very different structure appears. The integrand contains the determinant of a $(n-3)\times (n-3)$ matrix and a `Jacobian'. We prove that the determinant is a polynomial in the coefficients of the maps and contains ${\rm R}(\lambda )$ and ${\rm R}(\tilde\lambda )$ as factors. The equivalence of the two formulations then implies a dramatic simplification of the Jacobian part as it has to cancel the result of the polynomial division.


\end{abstract}

\bigskip
\bigskip

\end{titlepage}

\section{Introduction}

In 2003 Witten introduced a formulation for the S-matrix of ${\cal N}=4$ super Yang-Mills which can be thought of as  boosting the famous Parke-Taylor formula \cite{Parke:1986gb} for amplitudes of R-charge $k=2$ to amplitudes of arbitrary $R$-charge $k=d+1$ by using holomorphic maps of degree $d$ \cite{Witten:2003nn}. Shortly after, this formulation was studied and shown to satisfy important consistency conditions by Roiban, Spradlin and Volovich (RSV) \cite{Roiban:2004yf}. The Witten-RSV formulation for color-ordered amplitudes is very compact and it is given by
\be\nonumber
{\cal A}_{n,d}=\int d\Omega_n \int\prod_{\alpha =0}^d d^2\rho_\alpha\!\prod_{\alpha=0}^d\delta^2\!\left(\sum_{a=1}^n t_a\sigma_a^\alpha \tilde\lambda_a\right)\!\delta^4\!\left(\sum_{a=1}^n t_a\sigma_a^\alpha \tilde\eta_a\right)  \prod_{a=1}^n\delta^2\!\left(t_a\lambda(\sigma_a )\! -\! \lambda_a \right)
\ee
with
\be\label{hola}
d\Omega_n = \frac{1}{{\rm vol}(GL(2,{\mathbb{C}}))}\prod_{a=1}^n \frac{dt_a}{t_a}\frac{d\sigma_a}{(\sigma_a-\sigma_{a+1})} \quad {\rm and} \quad  \lambda (\sigma) = \sum_{\gamma=0}^d \rho_\gamma\sigma^\gamma.
\ee
In 2012, two analogous formulations for ${\cal N}=8$ supergravity were found almost simultaneously \cite{Cachazo:2012da,Cachazo:2012kg}. The first one was obtained by using the Kawai-Lewellen-Tye (KLT) relations \cite{Kawai:1985xq} which express gravity amplitudes in terms of products of two Yang-Mills amplitudes. Using the Witten-RSV formulation for the Yang-Mills amplitudes and some surprising orthogonality properties, the resulting object can be written as a single integral over holomorphic maps of degree $d$. The main ingredient in this formulation is a $n\times n$ matrix $\Psi$ of rank $n-3$ and its pseudo-determinant ${\rm det}'\Psi$ \cite{Cachazo:2012da} (this matrix is very reminiscent of that used by Hodges to construct amplitudes in the $k=2$ sector \cite{Hodges:2012ym} and thus makes this formulation a direct analog to the Witten-RSV formula). Important consistency checks such as parity invariance and soft limits were performed in \cite{Penante:2012hq}.

The second formulation was derived by combining two facts. The first is that gravity amplitudes break conformal invariance in twistor space in a very controlled manner as can be seen from BCFW recursion relations \cite{Mason:2009sa,ArkaniHamed:2009si}. The second is the realization that Hodges' formula \cite{Hodges:2012ym} for $k=2$ amplitudes in terms of a rank $n-3$ matrix can be naturally generalized if momentum conservation, the basic ingredient in Hodges' construction, is boosted from $2\times 2$ to $2\times k$ conditions (as exploited in a different context in \cite{Cachazo:2012uq}). In this way a matrix $\widetilde\Phi$ of rank ${\tilde d} = n-d-2$ was constructed. It turns out that another matrix was also needed; this time a matrix, $\Phi$, of rank $d$. Quite nicely, the roles of $\Phi$ and $\tilde\Phi$ are exchanged under parity. This formulation has been proven to reproduce all ${\cal N}=8$ supergravity amplitudes in \cite{Bullimore:2012cn,Cachazo:2012pz}. Moreover, a link representation analogous to that for Yang-Mills amplitudes \cite{ArkaniHamed:2009si,ArkaniHamed:2009dn,Spradlin:2009qr,Dolan:2009wf} was presented in \cite{He:2012er,Cachazo:2012pz}.

Very recently, Skinner has shown how to obtain the complete tree level S-matrix formulation in terms of $\Phi$ and $\widetilde\Phi$ from worldsheet correlation functions \cite{Skinner:2013xp}. Moreover, Skinner proved that in general the pseudo-determinant of the matrix $\Phi$, which enters in the formula, does not depend on worldsheet coordinates thus hinting that there must be a much simpler form for it. These exciting developments motivate the further examination of the two formulations mentioned above and the study of how they are related to each other.

In this work we show that when both formulations are written in a manifestly parity invariant form the determinants ${\rm det}'\Psi$, ${\rm det}'\widetilde\Phi$, and ${\rm det}'\Phi$ simplify substantially. In fact, we are able to prove that ${\rm det}'\Psi$ contains ${\rm det}'\widetilde\Phi$ and ${\rm det}'\Phi$ as factors. The key observation is that both ${\rm det}'\Phi$  and ${\rm det}'\widetilde\Phi$ are nothing but the resultants of maps from ${\mathbb CP}^1\to {\mathbb CP}^1$ of degrees $d$ and $\tilde d$ respectively. More explicitly, if $\sigma$ denotes an inhomogeneous coordinate of ${\mathbb CP}^1$ then the first map is given by a pair of polynomials in $\sigma$, i.e., $\lambda(\sigma) = (\lambda_1(\sigma),\lambda_2(\sigma ))$, each of degree $d$. The second map, $\tilde\lambda(\sigma) = (\tilde\lambda_1(\sigma),\tilde\lambda_2(\sigma ))$, is given a pair of polynomials of degree $\tilde d$. We denote the resultant of $\lambda_1(\sigma)$ and $\lambda_2(\sigma )$ with respect to $\sigma$ by ${\rm R}(\lambda)$. Likewise ${\rm R}(\tilde\lambda)$ denotes the resultant for the second map. Given that both maps are from the same ${\mathbb CP}^1$ it is more natural to talk about a single map $\sigma \to (\lambda(\sigma),\tilde\lambda(\sigma ))$ of bi-degree $(d,{\tilde d})$.

Let us present here the final result of the simplification which is derived in section 3. It is convenient to supersymmetrize the target space and turn it into ${\mathbb CP}^{1|4}\times {\mathbb CP}^{1|4}$. The maps then become
\be
{\mathbb L}(\sigma ) = \sum_{\alpha=0}^d {\mathbb M}_\alpha \sigma^\alpha, \quad \widetilde{\mathbb L}(\sigma ) = \sum_{\alpha=0}^{\tilde d} \widetilde{\mathbb M}_\alpha\sigma^{\alpha} \quad {\rm with} \quad {\mathbb M}_\alpha = \left(
                         \begin{array}{c}
                           \rho_\alpha \\
                           \chi_\alpha \\
                         \end{array}
                       \right), \quad \widetilde{\mathbb M}_\alpha =\left(
                         \begin{array}{c}
                           \tilde\rho_\alpha \\
                           \tilde\chi_\alpha \\
                         \end{array}
                       \right) ,
\ee
where $\rho_\alpha$'s are two component bosonic spinors while both $\chi_\alpha$'s and $\tilde\chi_\alpha$'s are four-component Grassmann vectors.

The external scattering data is usually given in the literature as $\{\lambda_a,\tilde\lambda_a,\tilde\eta_{a}\}$ with $\tilde\eta_{a}$ an eight-component Grassmann vector. In order to make parity invariance manifest it is more convenient to use
\be
{\mathbb L}_a = \left(\begin{array}{c}
                           \lambda_a \\
                           \eta^{\rm L}_a \\
                         \end{array}
                       \right) ,\quad\quad \widetilde{\mathbb L}_a = \left(\begin{array}{c}
                           \tilde\lambda_a \\
                           \tilde\eta^{\rm R}_a \\
                         \end{array}
                       \right)
\ee
with $\eta^{\rm L}$ and $\tilde\eta^{\rm R}$ four-component Grassmann vectors. 

Finally, it is useful to define a measure analogous to (\ref{hola})
\be
d\Omega_n = \frac{1}{{\rm vol}(GL(2,{\mathbb{C}}))}\prod_{a=1}^n d\sigma_a d t_a d{\tilde t}_a\delta\left(t_a{\tilde t_a} - \frac{1}{\prod_{b\neq a}(\sigma_a-\sigma_b)}\right)
\ee
to present the scattering amplitude of $n$ particles in the $k=d+1$ $R$-charge sector as
\vskip-0.1in
\be\nonumber
{\cal M}_{n,d}\! =\!\! \int\!\prod_{\alpha=0}^d d^{2|4}{\mathbb M}_\alpha \; {\rm R}(\lambda )\int\!\prod_{\beta=0}^{\tilde d} d^{2|4}\widetilde{\mathbb M}_\beta \; {\rm R}(\tilde\lambda )\int \!\! d\Omega_n \prod_{a=1}^n  \delta^{2|4}({\mathbb L}_a - t_a{\mathbb L}(\sigma_a))\; \delta^{2|4}(\widetilde{\mathbb L}_a - {\tilde  t}_a\widetilde{\mathbb L}(\sigma_a)).
\ee
One of the nicest features of this formula is that the only dependence on the marked points, $\sigma_a$'s, is through the delta function constraints and the measure $d\Omega_n$. While the original formulations are manifestly $SU(8)$ R-symmetric, this new form is only manifestly invariant under a $SU(4)\times SU(4)$ subgroup. 

This paper is organized as follows. In section 2 we illustrate the transformation from the usual formulations to the manifestly parity invariant one by applying it to ${\cal N}=4$ super Yang-Mills. In section 3 we transform both formulations of ${\cal N}=8$ supergravity amplitudes into a manifestly parity invariant form and obtained the formula in terms of resultants. In section 4, we prove that ${\rm det}'\Psi$ contains ${\rm det}'\widetilde\Phi$ and ${\rm det}'\Phi$ as factors and comment on the equivalence of the two formulations. Finally, in section 5 we end with discussions and future directions.

\section{Witten-RSV formulation of ${\cal N}=4$ SYM Amplitudes}

Before studying gravity amplitudes it is instructive to start by reviewing the Witten-RSV formulation of the S-matrix of ${\cal N}=4$ SYM \cite{Witten:2003nn,Roiban:2004yf}. The purpose is two-fold: firstly to establish notation and secondly to introduce the manifestly parity invariant formulation in a simpler set-up than gravity.

\subsection{The Witten-RSV Formula}

The formula for a color-ordered tree amplitude in the $k=d+1$ R-charge sector is given by
\be\nonumber
{\cal A}_{n,d}=\!\!\int d\Omega_n \!\prod_{\alpha =0}^d d^2\rho_\alpha\! \prod_{\alpha=0}^d\delta^2\!\left(\sum_{a=1}^n t_a\sigma_a^\alpha \tilde\lambda_a\right)\!\delta^4\!\left(\sum_{a=1}^n t_a\sigma_a^\alpha \tilde\eta_a\right)  \prod_{a=1}^n\delta^2\!\left(t_a\lambda(\sigma_a )\! -\! \lambda_a \right)
\ee 
with
\be
d\Omega_n = \frac{1}{{\rm vol}(GL(2,{\mathbb{C}}))}\prod_{a=1}^n \frac{dt_a}{t_a}\frac{d\sigma_a}{(\sigma_a-\sigma_{a+1})} \quad {\rm and} \quad \lambda (\sigma) = \sum_{\gamma=0}^d \rho_\gamma\sigma^\gamma.
\ee
The external data $\{\lambda_a\}$ and $\{\tilde\lambda_a \}$ enter this formula in very different ways. This is completely natural as superparticles have been chosen to be represented in $\tilde\eta$ space. While the asymmetry is actually welcomed in Yang-Mills as it makes the transformation to twistor space very simple, we will argue that in gravity it is more natural to select a more parity symmetric formulation. The key to finding the symmetric formulation was already given by Roiban, Spradlin and Volovich in version 2 of \cite{Roiban:2004yf} and by Witten in \cite{Witten:2004cp} while studying parity symmetry. In fact, Witten wrote ${\cal A}_{n,d}$ in essentially the form we need in \cite{Witten:2004cp}. The main identity that allows the transformation is
\be
\prod_{\alpha=0}^d\delta^2\left(\sum_{a=1}^n t_a\sigma_a^\alpha \tilde\lambda_a\right) = \left(V\prod_{b=1}^nt_b\right)^{-2}\int\prod_{\alpha=0}^{\tilde d}d^2\tilde\rho_\alpha \prod_{a=1}^n\delta^2\left({\tilde t}_a\tilde\lambda(\sigma_a ) - \tilde\lambda_a \right)
\ee
where
\be
{\tilde d} = n-d-2, \quad {\tilde t}_at_a = \frac{1}{\prod_{b\neq a}(\sigma_a-\sigma_b)}, \quad V =\prod_{a<b}(\sigma_a-\sigma_b), \quad {\rm and}\quad \tilde\lambda(\sigma) = \sum_{\gamma=0}^{\tilde d} \tilde\rho_\gamma\sigma^\gamma.
\ee
Also a transformation formula for the Grassmann variables is needed
\be\label{susy}
\int \prod_{a=1}^n d\eta_a^I \exp \left(\sum_{a=1}^n\eta_a^I\tilde\eta_{a,I}\right)\prod_{\alpha=0}^{\tilde d}\delta\left(\sum_{a=1}^n{\tilde t}_a\sigma_a^\alpha \eta_a^I \right) = \left(V\prod_{a=1}^n{\tilde t}_a\right)\prod_{\alpha=0}^d\delta\left(\sum_{a=1}^n t_a\sigma_a^\alpha \tilde\eta_{a,I} \right)
\ee
(There is no summation over the label $I$ in the argument of the exponential).

Let us consider the supersymmetric part in ${\cal A}_{n,d}$ and write it as
\be
\prod_{\alpha=0}^d\delta^4\!\left(\sum_{a=1}^n t_a\sigma_a^\alpha \tilde\eta_a\right) = \prod_{\alpha=0}^d\delta^2\!\left(\sum_{a=1}^n t_a\sigma_a^\alpha \tilde\eta_a^{L}\right)\prod_{\alpha=0}^d\delta^4\!\left(\sum_{a=1}^n t_a\sigma_a^\alpha \tilde\eta_a^{R}\right)
\ee
where each four component $\tilde\eta_a$ has been split into two two-component Grassmann vectors $\tilde\eta^L_a$ and $\tilde\eta^R_a$. Let us now switch from the $\tilde\eta^L$ representation to the $\eta^L$ one by Fourier transforming. And use (\ref{susy}) for all the delta functions containing $\tilde\eta_R$. After doing this we can represent each delta function in terms of an integral over auxiliary Grassmann variables $\chi$ and $\tilde\chi$. This allows us to use all integrations over $\tilde\eta^L$ and $\eta^R$ to produce new delta functions. The final result is
\be
\left(V\prod_{a=1}^n{\tilde t}_a\right)^{\! -2}\!\!\!\!\!\int\prod_{\alpha=0}^d d^2\chi_\alpha\prod_{a=1}^n\delta^2(\eta_a^L-t_a\sum_{\alpha=0}^d\chi_\alpha \sigma_a^\alpha )\int\prod_{\alpha=0}^{\tilde d} d^2\tilde\chi_\alpha\prod_{a=1}^n\delta^2(\tilde\eta_a^R-{\tilde t}_a\sum_{\alpha=0}^{\tilde d}\tilde\chi_\alpha \sigma_a^\alpha )
\ee
In analogy with the bosonic part, let us define Grassmann maps
\be
\eta (\sigma ) = \sum_{\alpha=0}^d\chi_\alpha \sigma^\alpha \quad \tilde\eta(\sigma )= \sum_{\alpha=0}^{\tilde d}\tilde\chi_\alpha \sigma^\alpha.
\ee
Note that the jacobian factors containing powers of $V$ cancel out since 
\be
\prod_{a=1}^nt_a{\tilde t}_a = V^{-2}.
\ee
Before combining all the ingredients it is convenient to introduce the parity conjugated version of each $t_a$ via
\be
1 = \int\prod_{a=1}^n d{\tilde t}_a \delta({\tilde t}_a - \frac{1}{t_a \prod_{b\neq a}(\sigma_a-\sigma_b)})
\ee
and a new measure
\be
d\Omega_n = \frac{1}{{\rm vol}(GL(2,{\mathbb{C}}))}\prod_{a=1}^n\frac{d\sigma_a}{\sigma_a-\sigma_{a+1}}dt_a d{\tilde t}_a\;\delta\left( t_a{\tilde t_a} - \frac{1}{\prod_{b\neq a}(\sigma_a-\sigma_b)}\right).
\ee
The final manifestly parity invariant formula is then
\be\nonumber
{\cal A}_{n,d} = \int d\Omega_n \int\prod_{\alpha=0}^d d^2\rho_\alpha d^2\chi_\alpha \prod_{a=1}^n\delta^2(\lambda_a - t_a\lambda(\sigma_a))\delta^2(\eta_a^L -t_a\eta(\sigma_a)) & \\ \nonumber  \int\prod_{\alpha=0}^{\tilde d} d^2\tilde\rho_\alpha d^2\tilde\chi_\alpha \prod_{a=1}^n\delta^2(\tilde\lambda_a - {\tilde t}_a\tilde\lambda(\sigma_a))\delta^2(\tilde\eta_a^R -{\tilde t}_a\tilde\eta(\sigma_a)). & 
\ee

\section{${\cal N}=8$ Supergravity Formulas}

In this section we show how to write the two formulations for the S-matrix of ${\cal N}=8$ supergravity, which are analogous to the Witten-RSV formula, in a manifestly parity invariant form and show how they greatly simplify.
 
\subsection{CS Formulation}

Let us start with the formulation obtained by the author and Skinner in \cite{Cachazo:2012kg}. The reason to start with this formulation is that the integrand will easily simplify.

The formula requires the introduction of two singular $n\times n$ matrices $\Phi$ and $\tilde\Phi$ of ranks $d$ and $\tilde d$ respectively. Let us concentrate on the first of the two. The matrix $\Phi$ is defined as follows
\be\nonumber
\Phi_{ab} = \frac{\langle \lambda(\sigma_a)\;\lambda(\sigma_b)\rangle}{(\sigma_a-\sigma_b)} \qquad {\rm for }\qquad a\neq b\qquad\qquad\qquad\qquad ~\; & \\
\Phi_{aa} = -\sum_{c\neq a}^n \frac{\langle\lambda(\sigma_a)\;\lambda(\sigma_c)\rangle}{(\sigma_a-\sigma_c)}\prod_{\alpha =0}^{\tilde d}\frac{(\sigma_c-\sigma_{m_\alpha})}{(\sigma_a-\sigma_{m_\alpha})}\frac{\prod_{k\neq a} (\sigma_a-\sigma_k)}{\prod_{l\neq c} (\sigma_c-\sigma_l)}.  &
\ee
where $\sigma_{m_\alpha}$ are reference points. 
%

The matrix $\Phi$ has rank $d$ and the way it enters in the gravity formula is through a determinant obtained after removing $n-d$ rows and columns. Let the set of rows and columns that remain be $\{r_1,\ldots,r_d \}$ and $\{c_1,\ldots ,c_d \}$ respectively. The first factor in the integrand is then
\be\label{detprime}
{\rm det}'(\Phi) = \frac{|\Phi_{\rm red}|}{|r_1\cdots r_d||c_1\cdots c_d|}
\ee  
where e.g. $|r_1\cdots r_d|$ is a Vandermonde determinant defined as the product of all $(\sigma_{r_a}-\sigma_{r_b})$ with $a<b$. In this form is not obvious that ${\rm det}'(\Phi)$ is independent of the choices made and hence permutation invariant. Both these facts were proven in \cite{Cachazo:2012kg}. Surprisingly, more is true; the diagonal terms can be simplified in such a way that all reference points are removed (This fact was first noticed by Mason and recently used by Skinner in \cite{Skinner:2013xp}).

Introducing the notation 
$$\lambda'(\sigma) = \frac{\d}{\d \sigma}\lambda(\sigma)$$
the simplification reads 
\be\label{oui}
-\sum_{c\neq a}^n \frac{\langle \lambda(\sigma_a)~\lambda(\sigma_c)\rangle}{(\sigma_a-\sigma_c)}\prod_{\alpha =0}^{\tilde d}\frac{(\sigma_c-\sigma_{m_\alpha})}{(\sigma_a-\sigma_{m_\alpha})}\frac{\prod_{k\neq a} (\sigma_a-\sigma_k)}{\prod_{l\neq c} (\sigma_c-\sigma_l)} = -\langle \lambda(\sigma_a)\;\lambda'(\sigma_a)\rangle.
\ee
The proof is actually quite simple. Construct a function of one complex variable $z$ and perform a contour integral around infinity
\be\label{contour}
\frac{\prod_{k\neq a}(\sigma_a-\sigma_k)}{\prod_{m=0}^{\tilde d}(\sigma_a-\sigma_{m_\alpha})}\oint\frac{dz}{(z-\sigma_a)^2}\frac{\langle \lambda(\sigma_a)\;\lambda(z)\rangle}{\prod_{b\neq a}(z-\sigma_b)}\prod_{\alpha=0}^{\tilde d}(z-\sigma_{m_\alpha}).
\ee
It is easy to check that the integrand goes as $1/z^2$ as $z\to \infty$ and therefore the integral vanishes. Deforming the contour one finds that (\ref{oui}) is just the fact that the sum over all residues at finite $z$'s vanishes. 

The matrix $\Phi$ is then given by
\be
\Phi_{ab} = \frac{\langle \lambda(\sigma_a)\;\lambda(\sigma_b)\rangle}{(\sigma_a-\sigma_b)} \quad {\rm for }\quad a\neq b\qquad {\rm and} \qquad \Phi_{aa} = -\langle \lambda(\sigma_a)\;\lambda'(\sigma_a)\rangle. 
\ee
%
%
A very important observation made in \cite{Skinner:2013xp} is that ${\rm det}'(\widehat\Phi)$ is only a function of $\rho_\alpha$, i.e., it is independent of all $\sigma_a$ variables. The proof given in \cite{Skinner:2013xp} is to observe that as a rational function in $\sigma_a$, ${\rm det}'(\widehat\Phi)$ has degree zero and no poles and hence it is a constant. Here let us give another proof which, as byproduct, produces the explicit form which is independent of the $\sigma_a$ variables.

Consider the function
\be\label{resultant1}
\frac{\langle \lambda(x)\; \lambda(y)\rangle}{x-y} = \sum_{\alpha,\beta =0}^{d-1}c_{\alpha,\beta}x^\alpha y^\beta
\ee
where $\{ c_{\alpha,\beta}\}$ are clearly polynomials in the $SL(2,\mathbb{C})$ invariants $\langle \rho_\gamma~\rho_\delta\rangle$. (Recall that the map is defined as
\be
\lambda(\sigma ) = \sum_{\alpha =0}^d\rho_\alpha \sigma^\alpha
\ee 
where $\rho_\alpha$ are two-component spinors.)

The key observation is to note that the right hand side of (\ref{resultant1}) defines a bilinear form ${\cal C}$ whose entries are the coefficients $c_{\alpha,\beta}$, i.e.
\be
\left(
  \begin{array}{cccc}
    1 & x & \ldots & x^{d-1} \\
  \end{array}
\right)\left(
         \begin{array}{cccc}
           c_{00} & c_{01} & \ldots & c_{0,d-1} \\
           c_{10} & c_{11} & \ldots & c_{1,d-1} \\
           \vdots & \vdots  & \ddots & \vdots \\
           c_{0,d-1} & c_{1,d-1} & \ldots & c_{d-1,d-1} \\
         \end{array}
       \right)
\left(
          \begin{array}{c}
            1 \\
            y \\
            \vdots \\
            y^{d-1} \\
          \end{array}
        \right).
\ee

Let us construct a basis for the space $\mathbb{C}^d$ as follows
\be
{\cal T} = \{\left(
  \begin{array}{c}
    1 \\
    \sigma_1 \\
    \vdots \\
    \sigma_1^{d-1} \\
  \end{array}
\right) ,\left(
  \begin{array}{c}
    1 \\
    \sigma_2 \\
    \vdots \\
    \sigma_2^{d-1} \\
  \end{array}
\right),\ldots , \left(
  \begin{array}{c}
    1 \\
    \sigma_d \\
    \vdots \\
    \sigma_d^{d-1} \\
  \end{array}
\right) \}.
\ee

It is easy to show that the bilinear ${\cal C}$ when expressed in the basis ${\cal T}$ is nothing but the matrix obtained from $\Phi$ by removing the last $n-d$ rows and the last $n-d$ columns! Let's denote such matrix as $\Phi_{\rm red}$.

Given that this is just a change of basis, the determinant of this matrix is nothing but the determinant of ${\cal C}$ times the determinant of the change of basis squared. Explicitly,
\be
{\rm det}(\Phi_{\rm red}) = {\rm det\; {\cal C}}\; \left| \left(
                                                 \begin{array}{cccc}
                                                   1 & 1 & \ldots & 1 \\
                                                   \sigma_1 & \sigma_2 & \ldots & \sigma_d \\
                                                   \vdots & \vdots & \ddots & \vdots \\
                                                   \sigma_1^{d-1} & \sigma_2^{d-1} & \vdots & \sigma_{d}^{d-1} \\
                                                 \end{array}
                                               \right)
 \right|^2.
\ee
The determinant on the right hand side involving $\sigma_a$'s is nothing but the Vandermonde determinant $|12\ldots d|$ that appears in (\ref{detprime}) and therefore we conclude that
\be
{\rm det}'(\Phi) =  {\rm det \; {\cal C}} 
\ee

Luckily, the matrix ${\cal C}$ defined in (\ref{resultant1}) is a famous one in elimination theory and it is known as the Bezout-Cayley matrix \cite{Book}. The entries are given by
\be
c_{AB} = \sum_{\alpha=0}^{{\rm Min}(A,B)}\langle \rho_{\alpha}~\rho_{A+B+1-\alpha}\rangle
\ee
with $A$ and $B$ ranging in the set $\{ 0,1,\ldots ,d-1\}$. 

Let us give some examples.

{\it Degree one:}

\be
{\cal C} = \langle \rho_0~\rho_1\rangle
\ee

{\it Degree two:}

\be
{\cal C} = \left(
             \begin{array}{cc}
               \langle \rho_0~\rho_1\rangle & \langle \rho_0~\rho_2\rangle \\
               \langle \rho_0~\rho_2\rangle & \langle \rho_1~\rho_2\rangle \\
             \end{array}
           \right)
\ee

{\it Degree three:}

\be
{\cal C} = \left(
             \begin{array}{ccc}
               \langle \rho_0~\rho_1\rangle & \langle \rho_0~\rho_2\rangle & \langle \rho_0~\rho_3\rangle \\
               \langle \rho_0~\rho_2\rangle & \langle \rho_0~\rho_3\rangle + \langle \rho_1~\rho_2\rangle & \langle \rho_1~\rho_3\rangle \\
               \langle \rho_0~\rho_3\rangle & \langle \rho_1~\rho_3\rangle & \langle \rho_2~\rho_3\rangle \\
             \end{array}
           \right)
\ee

{\it Degree four:}

\be
{\cal C} = \left(
             \begin{array}{cccc}
               \langle \rho_0~\rho_1\rangle & \langle \rho_0~\rho_2\rangle & \langle \rho_0~\rho_3\rangle & \langle \rho_0~\rho_4\rangle \\
               \langle \rho_0~\rho_2\rangle & \langle \rho_0~\rho_3\rangle+\langle \rho_1~\rho_2\rangle & \langle \rho_0~\rho_4\rangle + \langle \rho_1~\rho_3\rangle & \langle \rho_1~\rho_4\rangle \\
               \langle \rho_0~\rho_3\rangle & \langle \rho_0~\rho_4\rangle + \langle \rho_1~\rho_3\rangle & \langle \rho_1~\rho_4\rangle + \langle \rho_2~\rho_3\rangle & \langle \rho_2~\rho_4\rangle \\
               \langle \rho_0~\rho_4\rangle & \langle \rho_1~\rho_4\rangle & \langle \rho_2~\rho_4\rangle & \langle \rho_3~\rho_4\rangle \\
             \end{array}
           \right)
\ee

The mathematical interpretation of this formula is quite natural in our set up. Note that the map from $\mathbb{CP}^1$ to $\mathbb{CP}^1$ is given by
\be
\sigma \mapsto \lambda(\sigma) =\sum_{\alpha=0}^d \rho_\alpha \sigma^\alpha .
\ee
More explicitly, we have two polynomials of degree $d$,
\be
\lambda_1(\sigma) =\sum_{\alpha=0}^d \rho_{\alpha,1} \sigma^\alpha, \quad \lambda_2(\sigma) =\sum_{\alpha=0}^d \rho_{\alpha,2} \sigma^\alpha.
\ee
One could expect that the amplitude does not receive contributions from regions in the moduli space where the map degenerates, i.e., from where both polynomials $\lambda_1(\sigma)$ and $\lambda_2(\sigma)$ can have a common root. The presence of such a common root is detected by computing the resultant of $\lambda_1(\sigma)$ and $\lambda_2(\sigma)$  viewed as univariate polynomials in $\sigma$. Let us denote such resultant as ${\rm R}(\lambda_1,\lambda_2)$. In order to make the fact that the resultant is invariant under $SL(2,\mathbb{C})$ transformations acting on $\{\lambda_1(\sigma),\lambda_2(\sigma)\}$ we will denote it as $R(\lambda )$. It is a classic result that (see for example \cite{Book})
\be
{\rm R}(\lambda) = {\rm det}\; {\cal C}.
\ee

Now it should be clear why the manifestly parity invariant form is very convenient for gravity. This is apparent when considering the second matrix needed in the gravity formula, i.e., $\tilde\Phi$. As the reader can anticipate due to the notation, $\tilde\Phi$ is nothing but the parity conjugated version of $\Phi$. While in the original formulation both matrices are quite different, in the manifestly parity invariant form they have identical structures. More explicitly, if we define
\be\label{resultant2}
\frac{[ \tilde\lambda(x)\;\tilde\lambda(y)]}{x-y} = \sum_{\alpha,\beta =0}^{{\tilde d}-1}{\tilde c}_{\alpha,\beta}x^\alpha y^\beta
\ee
then the final ingredient becomes 
\be
{\rm det}'(\tilde\Phi) =  {\rm det \; \widetilde{\cal C}} = {\rm R}(\tilde\lambda ).
\ee
Let us introduce a measure as in the case of Yang-Mills amplitudes
\be
d\Omega_{n} = \frac{1}{{\rm vol}(GL(2,{\mathbb C}))}\prod_{a=1}^nd\sigma_a dt_a d\tilde t_a \prod_{a=1}^n\delta\left(t_a{\tilde t_a} - \frac{1}{\prod_{b\neq a}(\sigma_a-\sigma_b)}\right).
\ee
Now we are ready to write the formula for the gravity amplitude as 
\be\nonumber
{\cal M}_{n,d} = \int d\Omega_{n} \int\prod_{\alpha=0}^d d^2\rho_\alpha d^4\chi_\alpha\prod_{a=1}^n \delta^2(\lambda_a - t_a\lambda(\sigma_a))\delta^4(\eta_a^L -t_a\chi(\sigma_a)){\rm R}(\lambda ) & \\ \nonumber \int\prod_{\alpha=0}^{\tilde d} d^2\tilde\rho_\alpha d^4\tilde\chi_\alpha \prod_{a=1}^n  \delta^2(\tilde\lambda_a - {\tilde t}_a\tilde\lambda(\sigma_a))\delta^4(\tilde\eta_a^R -{\tilde t}_a\tilde\chi(\sigma_a)){\rm R}(\tilde\lambda ). &
\ee
This formula suggest that the maps should be supersymmetrized and be thought of as maps from ${\mathbb CP}^1$ to ${\mathbb CP}^{1|4}\times {\mathbb CP}^{1|4}$. As anticipated in the introduction, the maps become
\be
{\mathbb L}(\sigma ) = \sum_{\alpha=0}^d {\mathbb M}_\alpha \sigma^\alpha, \quad \widetilde{\mathbb L}(\sigma ) = \sum_{\alpha=0}^{\tilde d} \widetilde{\mathbb M}_\alpha\sigma^{\alpha} \quad {\rm with} \quad {\mathbb M}_\alpha = \left(
                         \begin{array}{c}
                           \rho_\alpha \\
                           \chi_\alpha \\
                         \end{array}
                       \right), \quad \widetilde{\mathbb M}_\alpha =\left(
                         \begin{array}{c}
                           \tilde\rho_\alpha \\
                           \tilde\chi_\alpha \\
                         \end{array}
                       \right) .
\ee
In the usual formulations of ${\cal N}=8$ supergravity scattering amplitudes, the external scattering data is given as $\{\lambda_a,\tilde\lambda_a,\tilde\eta_a\}$ with $\tilde\eta_a$ an eight-component Grassmann vector. Instead, in our current set up is it more natural to introduce
\be
{\mathbb L}_a = \left(\begin{array}{c}
                           \lambda_a \\
                           \eta^{\rm L}_a \\
                         \end{array}
                       \right) ,\quad\quad \widetilde{\mathbb L}_a = \left(\begin{array}{c}
                           \tilde\lambda_a \\
                           \tilde\eta^{\rm R}_a \\
                         \end{array}
                       \right)
\ee
with $\eta^{\rm L}$ and $\tilde\eta^{\rm R}$ four-component Grassmann vectors. 

Finally we find a very compact formula for the amplitude of $n$ particles in the $k=d+1$ $R$-charge sector as
\vskip-0.1in
\be\nonumber
{\cal M}_{n,d}\! =\!\! \int\!\prod_{\alpha=0}^d d^{2|4}{\mathbb M}_\alpha \; {\rm R}(\lambda )\int\!\prod_{\beta=0}^{\tilde d} d^{2|4}\widetilde{\mathbb M}_\beta \; {\rm R}(\tilde\lambda )\int \!\! d\Omega_n \prod_{a=1}^n  \delta^{2|4}({\mathbb L}_a - t_a{\mathbb L}(\sigma_a))\; \delta^{2|4}(\widetilde{\mathbb L}_a - {\tilde  t}_a\widetilde{\mathbb L}(\sigma_a)).
\ee

Note that the only dependence on the marked points $\sigma_a$ is through the delta function constrains and the measure $d\Omega_n$. Moreover, the integral almost factors into two sectors; left (L) and right (R). Each sector is manifestly ${\cal N}=4$ supersymmetric. This means that manifest $SU(8)$ R-symmetry, which was present in the original formulation, is lost in favor of manifest parity invariance.

For completeness, let us also write the same formula as an integral over the Grassmannian $G(2,n)$. This is the analog to the Yang-Mills construction of \cite{ArkaniHamed:2009dg}. This also serves as a good introduction to the discussion in the next subsection. In order to do so one has to integrate out all ${\tilde t}_a$'s and use the change of variables
\be
\left(
  \begin{array}{c}
    \sigma_{1}^{(a)} \\
    \sigma_{2}^{(a)} \\
  \end{array} 
\right) = \left(
  \begin{array}{c}
    t^{1/d}_a \\
    t^{1/d}_a\sigma_{a} \\
  \end{array}
\right). 
\ee
A straightforward computation (which involves a rescaling of the $\widetilde{\mathbb M}_\beta$) gives rise to 
\be\nonumber
{\cal M}_{n,d}\! = \!\!\int\!\prod_{\alpha=0}^d d^{2|4}{\mathbb M}_\alpha \; {\rm R}(\lambda )\int\!\prod_{\beta=0}^{\tilde d} d^{2|4}\widetilde{\mathbb M}_\beta \; {\rm R}(\tilde\lambda )\int \frac{d^{2n}\sigma}{{\rm vol}(GL(2,{\mathbb C}))}\prod_{a=1}^n  \delta^{2|4}({\mathbb L}_a - \sum_{\alpha=0}^d {\mathbb M}_\alpha C^{V}_{\alpha, a}(\sigma))& \\ \prod_{a=1}^n\delta^{2|4}(\widetilde{\mathbb L}_a - \sum_{\beta=0}^{\tilde d}\widetilde{\mathbb M}_\alpha {\widetilde C}^{V}_{\beta, a}(\sigma)). \qquad \qquad\qquad\qquad\qquad\qquad\qquad \qquad\qquad \qquad\qquad ~~ &
\ee
with
\be
C^{V}_{\alpha, a}(\sigma) = (\sigma_{1}^{(a)})^{d-\alpha}(\sigma_{2}^{(a)})^\alpha, \qquad
{\widetilde C}^{V}_{\beta, a}(\sigma) = \frac{(\sigma_{1}^{(a)})^{{\tilde d}-\beta}(\sigma_{2}^{(a)})^\beta}{\prod_{b\neq a}(a\;b)}.
\ee
and $(a\; b)$ denoting the Plucker coordinates of $G(2,n)$.

\subsection{CG Formulation} 
 
The other formulation for the S-matrix of ${\cal N}=8$ supergravity is very different from the previous one and was obtained by the author and Geyer in \cite{Cachazo:2012da}. In this formulation one also has two singular matrices. However, the two matrices are not mapped into each other under parity. Moreover, the integrand is the ratio of the two pseudo-determinants rather than the product.

The first matrix is very analogous to Hodges' formula for MHV amplitudes \cite{Hodges:2012ym}. The original form presented in \cite{Cachazo:2012da} is given in terms of an integral over $G(2,n)$. The matrix, $\Psi_{ab}$, is then defined as
\be\nonumber
\Psi_{ab} = \frac{\langle \lambda_a\;\lambda_b\rangle [\tilde\lambda_a\;\tilde\lambda_b]}{(a\; b)^2} \qquad\qquad\qquad\qquad ~~\; a\neq b, & \\ \Psi_{aa} = -\sum_{c\neq a} \frac{\langle \lambda_a\;\lambda_c\rangle [\tilde\lambda_a\;\tilde\lambda_c]}{(a\; c)^2}
                \frac{(c\; \ell)(c\; r)}{(a\;\ell)(a\; r)}
                  \qquad a=b.
\ee
%
%
In this form only the external data and $\sigma$ variables enter. In fact, only the Plucker coordinates $(a\; b)$ of $G(2,n)$ appear. We expect that this matrix should nicely simplify using the manifestly parity invariant formulation as $\Psi$ itself is symmetric under the exchange of $\lambda_a$ and $\tilde\lambda_a$.

The object that enters the integrand is obtained by removing three rows $\{ r_1,r_2,r_3\}$ and three columns $\{ c_1,c_2,c_3\}$ to get a reduced non-singular matrix $\Psi_{\rm red}$ and then compute
\be
{\rm det}' \Psi = \frac{|\Psi_{\rm red}|}{|r_1r_2r_3||c_1c_2c_3|}.
\ee
In order to simplify this form it is convenient to use the same change of variables as at the end of the previous subsection but in the opposite direction,
\be
\left(
  \begin{array}{c}
    \sigma^{(a)}_1 \\
    \sigma^{(a)}_2 \\
  \end{array}
\right) = \left(
  \begin{array}{c}
    t_a^{1/d} \\
    t_a^{1/d}\sigma_a \\
  \end{array}
\right).
\ee 

A simple but lengthy exercise shows that after performing the change of variables and using the support of the delta functions imposing that
\be
\lambda_a = t_a\lambda (\sigma_a),\quad \tilde\lambda_a = {\tilde t}_a\tilde\lambda (\sigma_a)
\ee 
one finds a new matrix $\widehat\Psi_{ab}$ with
\be\nonumber
\widehat\Psi_{ab} = \frac{\langle \lambda(\sigma_a)\;\lambda(\sigma_b)\rangle [ \tilde\lambda(\sigma_a)\;\tilde\lambda(\sigma_b)] }{(\sigma_a-\sigma_b)^2} \qquad\qquad\qquad\qquad\qquad\qquad\qquad\qquad\qquad\quad\; a\neq b \quad \\
               \widehat\Psi_{aa} = -\sum_{c=1}^n \frac{\langle \lambda(\sigma_a)\;\lambda(\sigma_c)\rangle [ \tilde\lambda(\sigma_a)\;\tilde\lambda(\sigma_c)]}{(\sigma_a-\sigma_c)^2}
                \frac{(\sigma_c-\sigma_\ell)(\sigma_c-\sigma_r)}{(\sigma_a-\sigma_\ell)(\sigma_a-\sigma_r)}
                \frac{\prod_{d\neq a}(\sigma_a-\sigma_d)}{\prod_{e\neq c}(\sigma_c-\sigma_e)} \quad\quad a=b \quad
\ee
%
%
%
and a somewhat surprising identity. The identity is easiest to describe by choosing, for example, both $\{ r_1,r_2,r_3\}$ and $\{ c_1,c_2,c_3\}$ to be $\{1,2,3\}$, then
\be
\frac{|\Psi_{\rm red}|}{|123|^2} =  \frac{|\widehat\Psi_{\rm red}|}{|45\cdots n|^2}\times \frac{1}{|12\cdots n|^2} 
\ee
where $\widehat\Psi_{\rm red}$ is also obtained from $\widehat\Psi$ by removing rows and columns $\{1,2,3\}$. In other words, the new denominator contains the Vandermonde of the rows and columns that remain.

%
%

At this point it is hard to miss the fact that the off-diagonal terms of $\widehat\Psi$ turn out to be the product of corresponding off-diagonal terms of $\Phi$ and $\tilde\Phi$ in the CS formulation. This suggest that the diagonal terms are also related in a simple way. In fact, the relation is exactly the same. More precisely,
%
%
%
%

\be\nonumber
 -\!\! \sum_{c=1}^n \frac{\langle \lambda(\sigma_a)\;\lambda(\sigma_c)\rangle [ \tilde\lambda(\sigma_a)\;\tilde\lambda(\sigma_c) ] }{(\sigma_a-\sigma_c)^2}
                \frac{(\sigma_c-\sigma_\ell)(\sigma_c-\sigma_r)}{(\sigma_a-\sigma_\ell)(\sigma_a-\sigma_r)}
                 \frac{\prod_{d\neq a}(\sigma_a-\sigma_d)}{\prod_{e\neq c}(\sigma_c-\sigma_e)}
\ee
turns out to be
$$\langle \lambda(\sigma_a)\;\lambda'(\sigma_a)\rangle [ \tilde\lambda(\sigma_a)\;\tilde\lambda'(\sigma_a)] $$
The proof is again straightforward. It only requires the generalization of the rational function introduced in (\ref{contour}) and the corresponding contour argument.


Quite nicely, ${\rm det}'\widehat\Psi$ defined as
\be
\frac{|\widehat\Psi_{\rm red}|}{|45\cdots n|^2}
\ee
also turns out to be $\sigma_a$ independent. In fact, we can follow exactly the same logic as explained in the previous section to manifestly remove all $\sigma_a$ dependence.

Consider the polynomial in $x$ and $y$,
\be
\frac{\langle \lambda(x)\;\lambda(y)\rangle [ \tilde\lambda(x)\;\tilde\lambda(y) ] }{(x-y)^2} =\sum_{\alpha,\beta=0}^{n-4}h_{\alpha,\beta} x^{\alpha}y^{\beta} 
\ee
and let ${\cal H}$ be the $(n-3)\times (n-3)$ matrix whose entries are $h_{\alpha,\beta}$. Once again, if we remove the first three columns and the first three rows of $\widehat\Psi$ one finds that $\widehat\Psi_{\rm red}$ is nothing but ${\cal H}$ is the basis of vectors obtained by setting $x$ in $(1,x,x^2,\ldots , x^{n-4})$ to one of the $n-3$ values $\{\sigma_4,\sigma_5,\ldots ,\sigma_{n}\}$ and therefore the matrix of the change of basis gives rise to the Vandermonde squared of the labels that remain and cancels the factor in the denominator of ${\rm det}'\Psi$. Summarizing,
\be
{\rm det}'{\widehat \Psi} = {\rm det}{\cal H}.
\ee
The entries of ${\cal H}$ are easily computed in terms of those of the Berzout-Cayley matrices ${\cal C}$ and $\widetilde{\cal C}$ found in the previous formulation. One has
\be
\sum_{\alpha,\beta=0}^{n-4}h_{\alpha,\beta} x^{\alpha}y^{\beta} = \left(\sum_{\alpha,\beta=0}^{d-1} c_{\alpha\beta}x^\alpha y^\beta \right)\left( \sum_{\tilde\alpha,\tilde\beta =0}^{{\tilde d}-1} {\tilde c}_{\tilde\alpha \tilde\beta}x^{\tilde\alpha} y^{\tilde\beta} \right).
\ee
This means that ${\cal H}$ can be thought of as the matrix convolution of ${\cal C}$ and $\widetilde{\cal C}$. This simple connection is the key to proving the relation between the two formulations. We postpone the proof to the next section. Instead, let us describe the final ingredient in the CG formulation.

The last piece turns out to be a Jacobian! In the current manifestly parity invariant form, the Jacobian is defined using the following set of equations
\be
{\cal E} = \{ t_1\lambda(\sigma_1)-\lambda_1 , \ldots ,t_n\lambda(\sigma_n)-\lambda_n, {\tilde t}_1\tilde\lambda(\sigma_1)-\tilde\lambda_1 ,\ldots ,{\tilde t}_n\tilde\lambda(\sigma_n)-\tilde\lambda_n \}
\ee
and variables
\be
{\cal V} = \{ \rho_1,\ldots ,\rho_n, \tilde\rho_1, \ldots ,\tilde\rho_n ,t_1,\ldots t_n,\sigma_1,\ldots ,\sigma_n \} .
\ee
Here we have assumed that ${\tilde t}_a$ has been solved for in terms of $t_a$ and $\sigma$'s. In other words, in these formulas ${\tilde t}_a={\tilde t}_a(t_a,\sigma's)$.

There is a total of $4n$ equations and $4n$ variables. Naively one would expect this system of equations to be non-singular. However, the straightforward Jacobian vanishes. In fact, the Jacobian matrix has corank four. This is a consequence of the fact that four of the delta functions actually constrain the external data $\lambda_a$ and $\tilde\lambda_a$ to satisfy momentum conservation. This is nicely matched by the fact that four of the variables should be ``gauge fixed" thanks to the $GL(2,\mathbb{C})$ invariance of the full integral.

The way to proceed is just as with the other singular matrices we have already encountered. Define a $4n\times 4n$ matrix ${\cal K}$ with entries
$${\cal K}_{IJ} =\frac{\d {\cal E}_I}{\d {\cal V}_J} $$
and compute ${\rm det}'{\cal K}$ by removing four rows and four columns. It is convenient to remove two of the spinor equations in $\lambda$, say the ones corresponding to labels $a$ and $b$ and the rows corresponding to the variables $\{t_a,t_b,\sigma_a,\sigma_b\}$. Then
\be
{\rm det}'{\cal K} = \frac{|{\cal K}_{\rm red}|}{|ab|^2[\tilde\lambda(\sigma_a)\;\tilde\lambda (\sigma_b)]^2}
\ee
where $|ab|=(\sigma_a-\sigma_b)$.

We will not prove it here but explicit computations for $d=2$ and $n=6,7$ suggest that the map and $\sigma$ dependence of ${\rm det}'{\cal K}$ also factor and give rise to
\be\label{maincon}
{\rm det}'{\cal K} = \frac{M(\lambda,\tilde\lambda)}{|12\cdots n|^2}
\ee
where $M(\lambda,\tilde\lambda)$ is a polynomial in the map coefficients. In the next section we present some explicit examples.

Combining the formulas of the various pseudo-determinants we conclude that the integrand of the CG gravity formula is 
\be
{\cal I} = \frac{{\rm det}'{\Psi} }{{\rm det}'{\cal K}} = \frac{{\rm det}{\cal H}(\lambda,\tilde\lambda)}{M(\lambda,\tilde\lambda)}.
\ee
In this formulation, the gravity amplitude can then be written as
\be
\nonumber
{\cal M}_{n,d}\! =\!\! \int\!\prod_{\alpha=0}^d d^{2|4}{\mathbb M}_\alpha\!\!\int\!\prod_{\alpha=0}^{\tilde d} d^{2|4}\widetilde{\mathbb M}_\alpha \frac{{\rm det}{\cal H}(\lambda,\tilde\lambda)}{M(\lambda,\tilde\lambda)}\!\int \!\! d\Omega_n\!\prod_{a=1}^n  \delta^{2|4}({\mathbb L}_a - t_a{\mathbb L}(\sigma_a))\! \prod_{a=1}^n\delta^{2|4}(\widetilde{\mathbb L}_a - {\tilde  t}_a\widetilde{\mathbb L}(\sigma_a)).
\ee
%
%
%
%
$$\quad $$

\section{Equivalence of the Two Gravity Formulations}

Having expressed the two formulations in manifestly parity invariant forms we can proceed to study their relation. Let us start this section by rewriting both formulas. The CS formulation is given by
\be\nonumber
{\cal M}_{n,d}\! =\!\! \int\!\prod_{\alpha=0}^d d^{2|4}{\mathbb M}_\alpha \; {\rm R}(\lambda )\int\!\prod_{\beta=0}^{\tilde d} d^{2|4}\widetilde{\mathbb M}_\beta \; {\rm R}(\tilde\lambda )\int \!\! d\Omega_n \prod_{a=1}^n  \delta^{2|4}({\mathbb L}_a - t_a{\mathbb L}(\sigma_a))\; \delta^{2|4}(\widetilde{\mathbb L}_a - {\tilde  t}_a\widetilde{\mathbb L}(\sigma_a))
\ee
while the CG formulation is
\be
\nonumber
{\cal M}_{n,d}\! =\!\! \int\!\prod_{\alpha=0}^d d^{2|4}{\mathbb M}_\alpha\!\!\int\!\prod_{\alpha=0}^{\tilde d} d^{2|4}\widetilde{\mathbb M}_\alpha \frac{{\rm det}{\cal H}(\lambda,\tilde\lambda)}{M(\lambda,\tilde\lambda)}\!\int \!\! d\Omega_n\!\prod_{a=1}^n  \delta^{2|4}({\mathbb L}_a - t_a{\mathbb L}(\sigma_a))\! \prod_{a=1}^n\delta^{2|4}(\widetilde{\mathbb L}_a - {\tilde  t}_a\widetilde{\mathbb L}(\sigma_a)).
\ee
It is clear that in order to prove the equivalence of the two formulations at the level of the integrand a very surprising identity must hold among polynomials in the map coefficients. Explicitly, it must be that
\be
{\rm det}{\cal H}(\lambda,\tilde\lambda) = R(\lambda) R(\tilde\lambda )M(\lambda,\tilde\lambda).
\ee
In this section we will prove that both $R(\lambda)$ and $R(\tilde\lambda )$ divide $ {\rm det}{\cal H}(\lambda,\tilde\lambda)$. This means that the result of the polynomial division, say ${\widehat M}(\lambda,\tilde\lambda)$, should then coincide with the map dependent part of the Jacobian, $M(\lambda,\tilde\lambda)$. This fact is illustrated in some examples at the end of this section. In fact, the equivalence of the two formulations implies that the conjectured form of the Jacobian, ${\rm det}'{\cal K}$ in (\ref{maincon}), must hold and predicts what $M(\lambda,\tilde\lambda)$ is. This could be established by showing that the CG formulation satisfies the BCFW recursion relations. We comment more on that in the next section.

\subsection{The Polynomial $R(\lambda) R(\tilde\lambda )$ Divides ${\rm det}{\cal H}(\lambda,\tilde\lambda)$}

In order to prove that ${\rm det}{\cal H}$ has both $R(\lambda)$ and $R(\tilde\lambda)$ as factors it is simplest to start with the definition
\be
\sum_{\alpha,\beta=0}^{n-4}h_{\alpha,\beta} x^{\alpha}y^{\beta} = \left(\sum_{\alpha,\beta=0}^{d-1} c_{\alpha\beta}x^\alpha y^\beta \right)\left( \sum_{\tilde\alpha,\tilde\beta =0}^{{\tilde d}-1} {\tilde c}_{\tilde\alpha \tilde\beta}x^{\tilde\alpha} y^{\tilde\beta} \right).
\ee
Now assume that $R(\lambda)=0$, this means that there exists an $x_*$ (the common root of $\lambda_1(\sigma)$ and $\lambda_2(\sigma)$) such that $\sum_{\alpha=0}^{d-1}c_{\alpha\beta}x_*^\alpha = 0$. Therefore,
\be
\sum_{\alpha,\beta=0}^{n-4}h_{\alpha,\beta} x_*^{\alpha}y^{\beta} = 0
\ee 
for any $y$. But the vectors made from $(1,y,\ldots ,y^{d+{\tilde d}-2})$ by choosing $d+{\tilde d}-1$ generic $y$'s form a basis of the space $\mathbb{C}^{d+{\tilde d}-1}$ and therefore,
\be
\sum_{\alpha=0}^{n-4}h_{\alpha,\beta} x_*^{\alpha} =0.
\ee
This means that ${\cal H}$ has a null eigenvector and therefore ${\rm det}{\cal H}=0$ whenever ${\rm R}(\lambda)=0$. Of course, this proves that ${\rm R}(\lambda)$ divides $({\rm det}{\cal H})^r$ for some $r>0$. Clearly, $r=1$ if ${\rm R}(\lambda)$ does not have any double (or higher) roots. But since ${\rm R}(\lambda)$ is generic, it only has simple roots and $r=1$. The same logic applies to ${\rm R}(\tilde\lambda )$ and we have proven the desired result.

\subsection{Examples and the Jacobian}

Let us now illustrate how the assumption made on the Jacobian, ${\rm det}'{\cal K}$, in the previous section works in two examples.

Before turning to the examples it is instructive to find as much information about the polynomial $M(\lambda ,\tilde\lambda )$ as possible. As a polynomial in $c_{\alpha\beta}$ and ${\tilde c}_{\alpha\beta}$ one has that the bi-degree of ${\rm det}{\cal H}$ is $(d+{\tilde d}+1,d+{\tilde d}+1)$ while that of ${\rm R}(\lambda)$ and ${\rm R}(\tilde\lambda )$ is $(d,0)$ and $(0,{\tilde d})$ respectively. This means that the bi-degree of $M(\lambda,\tilde\lambda )$ must be $({\tilde d}-1,d-1)$. It is interesting to note that $d$ and $\tilde d$ have ``switched places". 

Let us see this in action in the simplest non-trivial examples. In order to simplify the notation let us denote
\be\nonumber
\langle \rho_\alpha\; \rho_\beta\rangle = \langle \alpha \beta\rangle \quad {\rm and} \quad [\tilde\rho_\alpha\; \tilde\rho_\beta] = [\alpha \beta].
\ee
As the first example consider the case $d={\tilde d}=2$. In this case it is easy to compute ${\rm det}{\cal H}(\lambda,\tilde\lambda)$ explicitly as
\be
\left|
  \begin{array}{ccc}
    \langle 10\rangle [10] &\langle 20\rangle [10]+\langle 10\rangle [20] & \langle 20\rangle [20] \\
    \langle 20\rangle [10]+\langle 10\rangle [20] & \langle 21\rangle [10]+2\langle 20\rangle [20]+\langle 10\rangle [21] & \langle 21\rangle [20]+\langle 20\rangle [21] \\
    \langle 20\rangle [20] & \langle 21\rangle [20]+\langle 20\rangle [21] & \langle 21\rangle [21] \\
  \end{array}
\right|
\ee 
and check that it factors as 
\be
\left|
  \begin{array}{cc}
    \langle 01\rangle & \langle 02\rangle \\
    \langle 02\rangle & \langle 12\rangle \\
  \end{array}
\right|\times\left|
  \begin{array}{cc}
    \!\;[01] & \!\;[02] \\
    \!\;[02] & \!\;[12] \\
  \end{array}
\right|\times (2\langle 02\rangle[02]-\langle 12\rangle[01]-\langle 01\rangle[12]).
\ee
Clearly, the first two factors correspond to $R(\lambda)$ and $R(\tilde\lambda )$ respectively. The third factor has the correct degree to be $M(\lambda,\tilde\lambda)$ and therefore it must be that
\be
M(\lambda,\tilde\lambda) = 2\langle 02\rangle[02]-\langle 12\rangle[01]-\langle 01\rangle[12].
\ee 
Once can check that this is indeed the case by performing an explicit computation of ${\rm det}'{\cal K}$.  

The final example is $d=2$ and ${\tilde d}=3$. In this case one finds that ${\rm det}{\cal H}(\lambda,\tilde\lambda)$ is given by
\be
\left|
  \begin{array}{cccc}
    \langle 01\rangle[01] & \langle 02\rangle[01]+\langle 01\rangle[02] & \langle 02\rangle[02]+\langle 01\rangle[03] & \langle 02\rangle[03] \\
    \langle 02\rangle[01]+\langle 01\rangle[02] & h_{11} & h_{12} & \langle 12\rangle[03]+\langle 02\rangle[13] \\
    \langle 02\rangle[02]+\langle 01\rangle[03] & h_{12} & h_{22} & \langle 12\rangle[13]+\langle 02\rangle[23] \\
    \langle 02\rangle[03] & \langle 12\rangle[03]+\langle 02\rangle[13] & \langle 12\rangle[13]+\langle 02\rangle[23] & \langle 12\rangle[23] \\
  \end{array}
\right|
\ee 
with
\be\nonumber
h_{11} & = & \langle 12\rangle[01]+2\langle 02\rangle[02]+\langle 02\rangle([12]+[03]),\\ \nonumber
h_{12} & = & \langle 12\rangle[02]+\langle 02\rangle([12]+2[03])+\langle 01\rangle[13],\\ \nonumber
h_{22} & = & \langle 12\rangle([12]+[03])+2\langle 02\rangle[13]+\langle 01\rangle[23]. 
\ee

One can check that this determinant factorizes as
\be
\left|
  \begin{array}{cc}
    \langle 01\rangle & \langle 02\rangle \\
    \langle 02\rangle & \langle 12\rangle \\
  \end{array}
\right|\times \left|
             \begin{array}{ccc}
               [01] & [02] & [03] \\
               \left[ 02\right] & \left[ 03\right] + [12] & [13] \\
               \left[ 03\right] & [13] & [23] \\
             \end{array}
           \right|\times M
\ee
with 
$$M =-4\langle 02\rangle^2[03]-\langle 12\rangle^2[01]+2\langle 01\rangle\langle 02\rangle[13] -\langle 01\rangle^2[23]
+2\langle 02\rangle\langle 12\rangle[02] - \langle 01\rangle\langle 12\rangle[12] + \langle 01\rangle\langle 12\rangle[03] $$
which agrees perfectly with a direct computation of ${\rm det}'{\cal K}$.


\section{Discussion and Future Directions}

The manifestly parity invariant form of the S-matrix of ${\cal N}=8$ supergravity constructed by using maps of bi-degree $(d,{\tilde d})$ is very compact. The scattering of $n$ particles in the $k=d+1$ R-charge sector is computed by
\be\nonumber
{\cal M}_{n,d}\! =\!\! \int\!\prod_{\alpha=0}^d d^{2|4}{\mathbb M}_\alpha \; {\rm R}(\lambda )\int\!\prod_{\beta=0}^{\tilde d} d^{2|4}\widetilde{\mathbb M}_\beta \; {\rm R}(\tilde\lambda )\int \!\! d\Omega_n \prod_{a=1}^n  \delta^{2|4}({\mathbb L}_a - t_a{\mathbb L}(\sigma_a))\; \delta^{2|4}(\widetilde{\mathbb L}_a - {\tilde  t}_a\widetilde{\mathbb L}(\sigma_a)).
\ee
The resultants in this formula seem to be a natural part of the corresponding measures. Let us make this more precise. Consider for example the measure
\be
\prod_{\alpha=0}^d d^{2|4}{\mathbb M}_\alpha \; {\rm R}(\lambda )
\ee
and rescale the maps ${\mathbb M}_\alpha\to \gamma {\mathbb M}_\alpha$. Under this rescaling 
\be
\prod_{\alpha=0}^d d^{2|4}{\mathbb M}_\alpha \to \gamma^{-2(d+1)}\prod_{\alpha=0}^d d^{2|4}{\mathbb M}_\alpha \quad {\rm while}\quad
{\rm R}(\lambda )\to \gamma^{2d}{\rm R}(\lambda )
\ee
and therefore the rescaling of the full measure becomes $\gamma^{-2}$, i.e., it is degree independent! It would be interesting to find a more geometric understanding of this measure.

Finally, as mentioned in the previous section it would be very interesting to prove that the CG formulation satisfies BCFW recursion relations. Alternatively one could prove that ${\rm det}{\cal H}(\lambda,\tilde\lambda)$ vanishes whenever ${\rm det}'{\cal K}$ does. This should be possible as the jacobian, ${\rm det}'{\cal K}$, vanishing means that the system of equations becomes singular. This must be related to either degenerations of the underlying curve, i.e., when it becomes a nodal curve or by degenerations of the map itself.

Independently of the equivalence of the two formulations, it would be interesting to repeat the proof of factorization of ${\cal M}_{n,d}$ given in \cite{Cachazo:2012pz}. In that paper it was shown that if, say, $P_L=p_1+p_2+\ldots + p_{n_L-1}$ becomes $\lambda\tilde\lambda + s q $ with $s\ll 1$, then, e.g., the map $\lambda(\sigma )$ should be rescaled to become, in new coordinates,
\be
\lambda (\sigma ) = s\sum_{\alpha =0}^{d_L-1}\rho_\alpha \sigma^\alpha + \rho_{d_L}\sigma^{d_L} + s\sum_{\beta=d_L+1}^d\rho_\beta \sigma^\beta
\ee 
where $d_L$ is the degree of the `left' curve while $d-d_L$ is that of the `right' curve. This can also be written in a more suggestive form as
\be
\lambda(\sigma ) = \left( s\sum_{\alpha =0}^{d_L-1}\rho_\alpha \sigma^\alpha + \rho_{d_L}\sigma^{d_L}\right)\left( 1 + s\sum_{\beta=d_L+1}^d\hat\rho_\beta \sigma^{\beta-d_L} \right) + {\cal O}(s^2)
\ee
with $\hat\rho_\beta$ simply related to $\rho_\beta$ and $\rho_{d_L}$. Therefore we have $\lambda_1(\sigma ) = \lambda^{L}_1(\sigma)\lambda^{R}_1(\sigma)$ and $\lambda_2(\sigma ) = \lambda^{L}_2(\sigma)\lambda^{R}_2(\sigma)$. It is an obvious property of resultants that
\be\nonumber
{\rm R}(\lambda^{L}_1(\sigma)\lambda^{R}_1(\sigma),\lambda^{L}_2(\sigma)\lambda^{R}_2(\sigma)) = {\rm R}(\lambda^{L}_1(\sigma),\lambda^{L}_2(\sigma)){\rm R}(\lambda^{L}_1(\sigma),\lambda^{R}_2(\sigma)){\rm R}(\lambda^{R}_1(\sigma),\lambda^{L}_2(\sigma)){\rm R}(\lambda^{R}_1(\sigma),\lambda^{R}_2(\sigma)).
\ee
Noting that all cross terms give constants one finds 
\be
{\rm R}(\lambda^{L}_1(\sigma)\lambda^{R}_1(\sigma),\lambda^{L}_2(\sigma)\lambda^{R}_2(\sigma)) \sim {\rm R}(\lambda^{L}_1(\sigma),\lambda^{L}_2(\sigma)){\rm R}(\lambda^{R}_1(\sigma),\lambda^{R}_2(\sigma))
\ee
which is the correct behavior for the proof of factorization.

\section*{Acknowledgments}

 The author thanks A. Buchel, E. Casali, Y. Geyer, S. He, R. Myers, A. Sever, D. Skinner, P. Vieira, M. Wijnholt and E. Yuan for useful discussions. This research was supported in part by the NSERC of Canada and MEDT of Ontario.


\begin{thebibliography}{10}

\bibitem{Parke:1986gb}
  S.~J.~Parke and T.~R.~Taylor,
  ``An Amplitude for $n$ Gluon Scattering,''
  Phys.\ Rev.\ Lett.\  {\bf 56}, 2459 (1986).

\bibitem{Witten:2003nn}
E.~Witten, ``{Perturbative Gauge Theory as a String Theory in Twistor Space},''
  {\em Commun. Math. Phys.}, vol.~252, pp.~189--258, 2004.

\bibitem{Roiban:2004yf}
  R.~Roiban, M.~Spradlin and A.~Volovich,
  ``On the Tree Level S Matrix of Yang-Mills Theory,''
  Phys.\ Rev.\ D {\bf 70}, 026009 (2004)
  [hep-th/0403190].

\bibitem{Witten:2004cp}
  E.~Witten,
  ``Parity Invariance for Strings in Twistor Space,''
  Adv.\ Theor.\ Math.\ Phys.\  {\bf 8}, 779 (2004)
  [hep-th/0403199].

\bibitem{Cachazo:2012da}
  F.~Cachazo and Y.~Geyer,
  ``A `Twistor String' Inspired Formula For Tree-Level Scattering Amplitudes in N=8 SUGRA,''
  arXiv:1206.6511 [hep-th].



\bibitem{Cachazo:2012kg}
  F.~Cachazo and D.~Skinner,
  ``Gravity from Rational Curves,''
  arXiv:1207.0741 [hep-th].


\bibitem{Kawai:1985xq}
  H.~Kawai, D.~C.~Lewellen and S.~H.~H.~Tye,
  ``A Relation Between Tree Amplitudes of Closed and Open Strings,''
  Nucl.\ Phys.\ B {\bf 269}, 1 (1986).


\bibitem{Hodges:2012ym}
  A.~Hodges,
  ``A Simple Formula for Gravitational MHV Amplitudes,''
  arXiv:1204.1930 [hep-th].



\bibitem{Penante:2012hq}
  B.~Penante, S.~Rajabi and G.~Sizov,
  ``Parity Symmetry and Soft Limit for the Cachazo-Geyer Gravity Amplitude,''
  JHEP {\bf 1211}, 143 (2012)
  [arXiv:1207.4289 [hep-th]].



\bibitem{Mason:2009sa}
  L.~J.~Mason and D.~Skinner,
  ``Scattering Amplitudes and BCFW Recursion in Twistor Space,''
  JHEP {\bf 1001}, 064 (2010)
  [arXiv:0903.2083 [hep-th]].

\bibitem{ArkaniHamed:2009si}
  N.~Arkani-Hamed, F.~Cachazo, C.~Cheung and J.~Kaplan,
  ``The S-Matrix in Twistor Space,''
  JHEP {\bf 1003}, 110 (2010)
  [arXiv:0903.2110 [hep-th]].


\bibitem{Cachazo:2012uq}
  F.~Cachazo,
  ``Fundamental BCJ Relation in N=4 SYM From The Connected Formulation,''
  arXiv:1206.5970 [hep-th].

\bibitem{Bullimore:2012cn}
  M.~Bullimore,
  ``New Formulae for Gravity Amplitudes: Parity Invariance and Soft Limits,''
  arXiv:1207.3940 [hep-th].

\bibitem{Cachazo:2012pz}
  F.~Cachazo, L.~Mason and D.~Skinner,
  ``Gravity in Twistor Space and its Grassmannian Formulation,''
  arXiv:1207.4712 [hep-th].

\bibitem{ArkaniHamed:2009dn}
  N.~Arkani-Hamed, F.~Cachazo, C.~Cheung and J.~Kaplan,
  ``A Duality For The S Matrix,''
  JHEP {\bf 1003}, 020 (2010)
  [arXiv:0907.5418 [hep-th]].

\bibitem{Spradlin:2009qr}
  M.~Spradlin and A.~Volovich,
  ``From Twistor String Theory To Recursion Relations,''
  Phys.\ Rev.\ D {\bf 80}, 085022 (2009)
  [arXiv:0909.0229 [hep-th]].

\bibitem{Dolan:2009wf}
  L.~Dolan and P.~Goddard,
  ``Gluon Tree Amplitudes in Open Twistor String Theory,''
  JHEP {\bf 0912}, 032 (2009)
  [arXiv:0909.0499 [hep-th]].



\bibitem{He:2012er}
  S.~He,
  ``A Link Representation for Gravity Amplitudes,''
  arXiv:1207.4064 [hep-th].

\bibitem{Skinner:2013xp}
  D.~Skinner,
  ``Twistor Strings for N=8 Supergravity,''
  arXiv:1301.0868 [hep-th].

\bibitem{Book}
I. Gelfand, M. Kapranov and A. Zelevinsky,
``Discriminats, Resultants, and Multidimensional Determinants", Birkhauser Boston, 2008.

\bibitem{ArkaniHamed:2009dg}
  N.~Arkani-Hamed, J.~Bourjaily, F.~Cachazo and J.~Trnka,
  ``Unification of Residues and Grassmannian Dualities,''
  JHEP {\bf 1101}, 049 (2011)
  [arXiv:0912.4912 [hep-th]].


\end{thebibliography}

\end{document}